\documentclass[a4paper,12pt]{article}
\usepackage{etex}
\usepackage[T1]{fontenc}
\usepackage[letterpaper,tmargin=2.5cm,bmargin=2.5cm,rmargin=2.5cm,lmargin=2.5cm]{geometry}
\usepackage{amssymb,amsmath}
\usepackage{amsthm}
\usepackage{graphicx}
\usepackage{fancyhdr}
\usepackage{fancybox}
\usepackage{shadow}
\usepackage{empheq}
\usepackage{titlesec}
\usepackage{titletoc}
\usepackage{soul, color}
\usepackage{float}
\usepackage{shorttoc}
\usepackage{xcolor}
\usepackage{fancybox}
\usepackage{natbib}
\usepackage{dcolumn}
\usepackage{booktabs}
\usepackage{rotating}
\usepackage{bigstrut}
\usepackage{natbib}
\usepackage{aeguill}
\usepackage{array}
\usepackage{multirow}
\usepackage{tabu}
\everymath{\displaystyle}
\newtheorem{thm}{Theorem}

\usepackage{changepage}

\usepackage{hyperref}
\def\sym#1{\ifmmode^{#1}\else\(^{#1}\)\fi}
\usepackage{caption}
\usepackage{blindtext}

\usepackage{caption}
\usepackage{subcaption}
\usepackage{mathtools}

\usepackage{graphicx}

\usepackage{enumerate}

\usepackage{hyperref}
\hypersetup{backref, colorlinks=true, citecolor=blue!50!black}
\usepackage{graphicx}

\usepackage{titlesec}

\usepackage{graphicx}
\usepackage{setspace}
\begin{document}

\title{Spatial Differencing for Sample Selection Models with Unobserved Heterogeneity}
%Spatial Differencing with Caution: Sample Selection Models with Unobserved Heterogeneity
% Spatial Differencing for A Sample Selection Model with Unobserved Heterogeneity
\author{Alexander Klein\thanks{%
School of Economics, e-mail: A.Klein-474@kent.ac.uk } \hspace{2cm} Guy Tchuente\thanks{Corresponding author: School of Economics, University of Kent, e-mail: guytchuente@gmail.com, Address: Kennedy Building, Park Wood Road, Canterbury, Kent, CT2 7FS, Tel:+441227827249. } \\
%EndAName
University of Kent }
\date{ June 2020}
\maketitle

\abstract{This paper derives  identification, estimation and inference results using spatial differencing in sample selection models with unobserved heterogeneity. We show that under the assumption of smooth changes across space of the unobserved sub-location specific heterogeneities and inverse Mills ratio, key parameters of a sample selection model are identified. The smoothness of the sub-location specific heterogeneities implies a correlation in the outcomes. We assume that the correlation is restricted within a location or cluster and derive asymptotic results showing that as the number of independent clusters increases, the estimators are consistent and asymptotically normal.    We also propose a formula for standard error estimation. A Monte-Carlo experiment illustrates the small sample properties of our estimator. The application of our procedure to estimate the determinants of the municipality tax rate in Finland shows the importance of accounting for unobserved heterogeneity.  \\

 }

\section{Introduction}

In linear models, spatial differencing has been used to deal with unobserved omitted variables. The availability of geographical locations allowed empirical papers to take advantage of the spatial dimension of the data and control for various unobserved heterogeneity
(e.g. \cite{duranton2011assessing}, \cite{black1999better} or \cite{holmes1996effects}). In general, spatial differencing offers an identification strategy in the situations when researchers face cross-sectional data with unobserved heterogeneity \emph{and} lack suitable instrumental variables.  This paper extends spatial differencing to a model with sample selection.

%%A short review on the use of proxies to solve omitted variable bias start with Todd and Wolpin (2003).
For economists, the question of omitted variables is a serious concern in the context of nonexperimental data. The solution is straightforward when omitted variables are simply the result of not including all relevant variables for which data exist - we add such variables to the model to avoid the bias induced by their omission. When omitted variables are unobserved,
researchers have basically three options: they can use (i) proxies, (ii) instrumental variables, or (iii) differencing the data across time or space.

The proxies reduce the bias if they manage to capture the effect of the omitted variables such that what remains is uncorrelated with the error term.
 However, it is often the case that the proxies are imperfect, hence they may still be  related to the unobserved heterogeneity, or the error term if it turns out to be endogenous, or it could also be irrelevant after controlling for observed covariates.   In such cases, the inclusion of the proxy will not solve the bias problem, it may even exacerbate it.\footnote{ See \cite{todd2003specification} for details on the use of proxy and \cite{oster2019unobservable} for a rigorous treatment of the evaluation of the robustness to omitted variables.} The second solution - using a valid set of the instruments - may help alleviate the bias. However, as discussed in \cite{todd2003specification}, the ``quasi-experimental"  local average treatment effect (LATE) obtained in the instrumental variable model may not correspond to the \emph{ceteris paribus} and thus,
 may not correspond to the deep structural parameter of interest. Lastly, panel data sets allow researchers to control for the unobserved heterogeneity. They indeed help them to identify the causal effect when, for example, time constant unobserved heterogeneity might cause endogeneity problem and strong instruments satisfying exclusion restrictions cannot be found. However, there might be situations when such data sets are not available.

%What are the available solutions?
Our paper is a contribution to the literature identifying and estimating model parameters in the presence of unobserved omitted variables. We propose an identification strategy based on spatial differencing. As was discussed above, this approach has been
used in the context of linear regressions. However, little is known about its performance in non-linear models. We
extend spatial differencing into this direction, specifically to the case of cross-section data with sample-selection. We show that under
justifiable assumptions on the smoothness of the unobserved heterogeneity (i.e. spatially close individuals have similar unobserved heterogeneity and
the derivative of their Inverse Mill's ratio are similar), spatial differencing eliminates the
 unobserved effects even in the presence of a nonlinear element - in our case Mill's ratio. The parameters of interest of our sample selection model are estimated  using a standard two-step
approach of \cite{heckman1974shadow}  \cite{heckman1979sample}. We derive asymptotic properties
and propose a correction of standard errors accounting for the two-step nature of our estimation and spatial differencing. The asymptotic behavior of the estimator reveals important properties of spatial differencing that researchers would need to be cautious about. The new estimator and the standard errors correction are easy to implement.
%%This is the first example in the paper, I was wondering if we can propose a case where there is sample selection.

%\emph{Version 1:}The model of sample selection with unobserved spatial heterogeneity that we consider in this paper can be described as follows. Suppose we have a data set of firms located in cities and our variable of interest is the productivity of firms. It is known that the location of firms in cities is often driven by a self-selection process (e.g.citation to Combes et al 2012), hence the data generating process of our data is the driven by a sample selection. The productivity of firms depends, among other things, on the unobserved city-specific conditions. We can control for them with location (city)-specific dummies. However, there can be a considerable unobserved heterogeneity at a \emph{finer} spatial scale than the locations themselves which we will call sub-locations. Using our example, firms can be located in diverse neighborhoods within the cities. As a result, locations-specific dummies might not capture \emph{all} of that fine-scale unobserved heterogeneity and using sub-location (neighborhood) specific dummies is not really an option since we may quickly run out of degrees of freedom.
% Combes, Pierre‐Philippe, Gilles Duranton, Laurent Gobillon, Diego Puga, and Sébastien Roux. "The productivity advantages of large cities: Distinguishing agglomeration from firm selection." Econometrica 80, no. 6 (2012): 2543-2594. \emph{Version 2:}

The intuitions for the model of sample selection with unobserved spatial heterogeneity that we consider in this paper can be described as follows. Suppose we have a cross-sectional data on municipalities which are organized into larger geographical units called regions, and which have the authority to set the levels of local taxation. Municipality tax rates must be at least as high as the threshold set by a central government. As a result, municipalities self-select into those with the tax rates at the threshold and those above it. We are interested in what determines the municipalities' tax rates. The tax rate will depend on various socio-economic characteristics, such as  age composition of population and income, but also on amenities. These can depend on the region where the municipalities are located: for example, regions with natural landscapes might have different level and composition of amenities than regions without them. We can control for them with region-specific dummies. However there can be a considerable unobserved heterogeneity at municipality level. Controlling for that with municipality-specific dummies might not be an option, since we may quickly run out of degrees of freedom. Therefore, we face a problem of a self-selected cross-sectional sample with unobserved heterogeneity which we cannot fully control with dummies, and which has two spatial dimensions: high-level which we call locations (in our example regions), and low-level which we call sub-locations (in our example municipalities).

%AK Note: This is very important, and I would suggest that we describe it in a bit more details here. I would propose to define what is common-sub-location specific unobserved heterogeneity since it would not be clear to a first-time reader. Also, what it means to enjoy the benefits of independent sample and what is independent cluster design. I understand what you mean but a first-time reader might not. GT: Agreed I will think about a paragraph for that.

Spatial differencing will eliminate the sub-location specific unobserved heterogeneity. It will, at the same time, also induce a  correlation in the error terms. We take that correlation into account, and derive the asymptotic properties of our estimator using similar arguments to those present in the derivation of the asymptotic behavior of the clustered standard errors: the number of locations goes to infinity and the size of location is assumed random and bounded almost surely. We find that this result also extends to a linear model without sample selection. This has important implications that researchers need to be cautious about. Indeed, the consistency of estimator applied to the spatially differenced data requires \emph{(i)} a large number of locations, \emph{(ii)} a limited number of individuals in each location. Monte Carlo simulations also suggest that it would be better if the number of individuals in sub-locations were small as well. Before we continue, let's notice that locations in our model are equivalent to clusters and we use 'location' and 'cluster' interchangeably.

%Using our earlier example in which we had counties as \emph{locations}, villages in these counties as \emph{sub-locations}, and farms in these villages as \emph{neighbors}, the consistency requires a large number of counties and a small number of farms per village.

%%%%Literature review: omitted variable bias and Cluster asymptotic Oster(2019) and Hansen and Lee (2019)
Since our estimator is derived for a clustered sample with unobserved heterogeneity, this paper contributes to the literature on the selection correction in panel data.
In this literature, the main challenge is the presence of individual-specific unobserved heterogeneity in both the outcome and the selection equations.
The existing solutions are based on either a full model specification or on a differencing procedure. \cite{wooldridge1995selection} uses a Mundlak approach to specify the individual-specific unobserved heterogeneity in both equations. He also imposes a special functional form to the selection mechanism.
\cite{kyriazidou1997estimation}, on the other hand, does not impose strong restriction of the selection equation functional form and uses a nonparametric approach to difference out the unobserved fixed-effect.  \cite{rochina1999new} similarly relies on differencing to identify the parameters of the model, but she also imposes additional distributional assumptions to the selection equation.
Even if our problem has similarities with the selection correction in panel data model literature, the main difference is that we are observing a clustered cross-section.  In each cluster, there is a finer common sub-location specific unobserved heterogeneity shared by some individuals in that cluster.
This heterogeneity, however, is different from the cluster and individual-specific ones studied in panel data models and implies a different cluster asymptotic. Since the outcome of the individuals are not independent in our model, while it is in the panel data case, our asymptotic results are thus derived using a large number of cluster asymptotic with heterogenous, random and bounded cluster size.

The clustered dependence created by the finer sub-location specific unobserved heterogeneity relates our asymptotic discussion to the papers dealing with clustering at variance level (see \cite{wooldridge2010econometric} for a textbook treatment).  The asymptotic in that literature is derived using either a large or a fixed number of clusters. The fixed number of cluster leads to non-normal asymptotic and discussion about recent contributions can be found in \cite{hansen2019asymptotic}.
 A large number of cluster asymptotic was first derived by \cite{white1984asymptotic} and has been investigated by several authors allowing either fixed cluster size or heterogeneous cluster.
 Recent developments include \cite{hansen2019asymptotic} who propose conditions on the relation between the cluster sample sizes and the full sample in a regular asymptotic,
or \cite{djogbenou2019asymptotic} who derive asymptotic with varying cluster sizes and carry out a cluster wild bootstrap.
Our results complement this literature by extending the cluster asymptotic to a sample selection model.

We present an empirical application of our new estimator. We examine the determinants of tax rates across four hundred and eleven Finish municipalities spread across nineteen Finish regions. In 1999, Finish central government has decided to raise the lower bound of the tax rate municipalities could set from 0.2\% to 0.5\%. This has created a sample selection mechanism which resulted with more than half of the municipalities opting for 0.5\% while the rest charging higher tax rate. We use our spatial differencing estimator to control for unobserved municipalities effect which can be correlated with the error term, creating thus an endogeneity problem and rendering the standard sample selection estimator biased and inconsistent. Our results clearly show the presence of unobserved heterogeneity across municipalities, limitations of using only region dummies (nineteen in our case) to fully control for municipalities' unobserved heterogeneity, and the importance of spatial differencing to control for it.

The structure of the paper is as follows. First, we expand the spatial
differing method in the case of linear regression model to the case of
sample selection. Then we discuss identification assumptions, propose an
estimation procedure, and derive the estimator of the corrected
standard errors. Lastly, we conduct Monte Carlo simulations and present an empirical application of our estimator.

\section{Sample Selection Models with Spatial Correlation}

In many economic applications, we are interested in estimating the following
regression equation:
\begin{equation}
y_{ij}^{{}}=x_{ij}^{\prime }\delta +\gamma _{j}+\gamma _{j\alpha}+\varepsilon
_{ij}  \label{lin}
\end{equation}
where $x_{ij}^{\prime }$ is a vector of exogenous controls variables, $\gamma _{j}$ is location
fixed effect, $\gamma _{j\alpha}$ is a sub-location specific effect for sub-location $\alpha$
which is at a finer spatial scale than location $j$, and $\varepsilon _{ij}$
is the error term.\footnote{The  sub-location specific component, $\gamma_{j\alpha}$, is a simplification for $%
\gamma_{j\alpha_i}$. We are implicitly assuming that the  sub-location specific effects
are the same for all its individuals.}
%AK Note: we have two parentheses with assumptions here. I would suggest to leave them out since we discuss them later in the paper more rigorously);  and is the main source of unobserved source of endogeneity

Examples of application of this model can be found in the  estimation of the fertilizer effect on wheat crops yield in farms growing multiple crops, or the effect of local taxation on the growth of firms. The crop-yields depend on the soil quality of location $\gamma _{j}$ (e.g. a village), but also on the  sub-location specific soil composition (e.g. a farm in the village) \cite{collins2006soil}.  Similarly, the impact of local taxation on the growth of firms may vary by county but also sub-locations such as neighborhoods  as in \cite{duranton2011assessing}.
We can control for $\gamma _{j}$ with location dummy variables. However, they
might not be enough to capture all unobserved heterogeneity related to
location $j$ as there can be considerable heterogeneity at finer spatial scale of sub-locations: using the example above, the firms are located in various
neighborhoods $\alpha$ which are sub-locations of location $j$. Furthermore, standard
location fixed effect $\gamma _{j}$ relies upon an arbitrary specification
of the comparison neighborhood group, as pointed out by \cite{gibbons2003valuing}, making it an imperfect control
for  sub-location specific effect $\gamma _{j\alpha}$. If $\gamma _{j\alpha}$ is correlated
with $x_{ij}$, OLS estimate of $\delta $ will be biased. In the absence of
suitable instrumental variables for $x_{ij}$, the spatial differencing
offers a solution by differencing out the unobserved  sub-location specific effects $%
\gamma _{j\alpha}$.

\cite{duranton2011assessing}, \cite{black1999better} or \cite%
{holmes1996effects} use spatial differencing in the case of linear models to
solve endogeneity problems arising from unobserved sub-location effect $\gamma _{j\alpha}$. They take advantage of the fact that for sufficiently small distances
between sub-locations, their specific effect $\gamma _{j\alpha}$ changes smoothly
across space, allowing thus to difference them out. This corresponds to the following assumption.\\

{\textbf{Assumption I1:} The sub-location specific unobservable effect is homogenous
in the a neighborhood of the individual ie $\Delta _{d}\gamma _{j\alpha}=0$ for $%
d $ small enough.\newline
}

In several economic models, in addition to the sub-location specific fixed effects $\gamma _{j\alpha}$%
, the outcome of interest is not observed for the selected sub-sample.  The selection can be the result of the decision of individuals or the researcher.  The presence of
sample selection introduces nonlinearity to the model (\ref{lin}).

We specify the model with sample selection as follows. Consider two latent
dependent variables $y^*_{1ij}$, and $y^*_{2ij}$ in a cross-section which follow a regular linear model for individual $i$ in a location $j$:

$y_{1ij}^{\ast }=z_{ij}^{\prime }\beta +\theta _{j\alpha}+\theta _{j}+\varepsilon
_{1ij}$ - selection equation,

$y_{2ij}^{\ast }=x_{ij}^{\prime }\delta +\gamma _{j\alpha}+\gamma
_{j}+\varepsilon _{2ij}$ - \ outcome equation.

\bigskip Individual error terms are $\varepsilon _{1ij}$ and $\varepsilon _{2ij}$ ; $\theta _{j\alpha}$ and $\gamma _{j\alpha}$ are sub-location specific effects for a sub-location $\alpha$  in location $j$, affecting the selection and the outcome equation respectively.
The exogenous characteristics $x_{ij}$ affect the outcome.  They could be correlated with $\gamma _{j\alpha}+\gamma
_{j}$ but not with $\varepsilon _{1ij}$ and $\varepsilon _{2ij}$. The variables $z_{ij}$  are exogenous variables determining selection, they can be a subset of $x_{ij}$. However, for identification purposes, some elements of $z_{ij}$  are assumed to be absent from $x_{ij}$.

%(AK note: the line below was probably a part of one sentence which was accidentally deleted?)

{\textbf{Assumption I2:} $\varepsilon _{1ij}$ and $\varepsilon _{2ij}$ are independent identically distributed normal random variables for all $i,j$.\\}
The outcome is modelled in the form of a truncated sample
selection model and is represented by equation (\ref{Model}).
\begin{equation}
y_{2ij}=\left\{
\begin{array}{c}
y_{2ij}^{\ast }\hspace{0.2cm}if\hspace{0.2cm}y_{1ij}^{\ast }>0 \\
-\hspace{0.2cm}if\hspace{0.2cm}y_{1ij}^{\ast }\leq 0%
\end{array}%
\right.  \label{Model}
\end{equation}

%(AK Note: I would suggest to reformat and maybe slightly rewrite the paragraph below. Maybe something along the lines: We can consistently estimate $\delta$  by Tobit regressions under the following assumptions: .... Could we move 'Condition 1, ... more to the right?)
Let us consider the following conditions.
\begin{enumerate}
\item[] \textbf{Condition 1:} $Cov[z_{ij}, \theta_{j\alpha}+\theta_j+\varepsilon_{1ij}]=0$; $z_{ij}$ is
exogenous

\item[] \textbf{Condition 2:} $Cov[x_{ij}, \gamma_{j\alpha}+\gamma_j+\varepsilon_{2ij}]=0$; $x_{ij}$ is
exogenous % \item  $Cov (\theta_{aj}, \gamma_{aj})=0$

\item[] \textbf{Condition 3:} and errors $(\varepsilon_{1ij}, \varepsilon_{2ij})$ satisfy $%
\varepsilon_{2ij}=\rho \times \varepsilon_{1ij} +v_{ij} $ with $%
\varepsilon_{1ij} \sim \mathcal{N}(0, 1)$ and independent of $v_{ij}$.
\end{enumerate}
It is possible to consistently  estimate  $\delta$  by Tobit regression under these three conditions.\footnote{%
Identification required an exclusion restriction ie a variable that affects $%
y^*_{1ij}$ but not $y^*_{2ij}$. Otherwise, identification  relies on the nonlinearity of
the inverse Mills ratio.} In most applications, the Condition 1 and 2 are unlikely to hold because there is a
possibility that, within a location, there could be a sub-location specific omitted variable affecting both the outcome and some observed
characteristics of interest. Thus, it is possible that $Cov[z_{ij}, \theta_{j\alpha}+\theta_j]\neq0$ and $Cov[x_{ij}, \gamma_{j\alpha}+\gamma_j]\neq 0$.  The standard way to deal with the correlation
between $x_{ij}$ and $\gamma _{j\alpha}$ would be to find a suitable instrument
for the $x_{ij}$ and run a IV Tobit or IV two-stage Heckit.

The very local nature of the sub-location specific effect means that it is not always evident to find a  variable correlated with $x_{ij}$
and uncorrelated with $\gamma_{j\alpha}$. %(AK note: I am not sure but I think some words are missing in the previous sentence?)
The exclusion restriction is likely to be violated and IV two-stage Heckit will yield inconsistent estimates for $\delta $.
Another option is to use the finer location fixed effect and estimate the model using classic Heckman two-stage procedure, but this will in practice lead to a proliferation of variable and lose of degrees of freedom.

\subsection{Identification via Spatial Differencing}

This section investigates the application of this spatial differencing technique to the
case of cross-section sample selection models.
We denote $\Delta _{d}$ to be a spatial difference operator. One example
is a pair-wise difference operator which takes the difference between each observation and another observation
located at distance less than $d$ from that observation. In a location $j$, with individual $i$ and $k$ who are neighbours. The pair-wise differencing of the variable $A$ is: $$\Delta_d A=A_{ij}-A_{kj}.$$

Another example is the difference between the individual outcome
and the average outcome of his/her neighbourhood $\mathcal{N}_{id}$. This operator is similar to the neighbourhood fixed effect operator, the difference being that the neighbourhoods can overlap. We call this operator the fixed-effect difference operator. Let $\mathcal{N}_{id}=\{ k,\hspace{0.2cm} in \hspace{0.2cm} neighbourhood \hspace{0.2cm}d  \}$, the sample size of $\mathcal{N}_{id}$ is $N_d$,  the differencing is given by: $$\Delta_{df} A=A_{ij}-\frac{1}{N_d}\sum_{k\in \mathcal{N}_{id}}A_{kj}.$$

 A further possibility is to use a kernel as in \cite{kyriazidou1997estimation} to weight neighbour in $\mathcal{N}_{id}$ according to how far they are, in term of observable characteristics. This operator is the kernel difference operator. $$\Delta_{dK} A=A_{ij}-\sum_{k\in \mathcal{N}_{id}}\psi(i,k) A_{kj}.$$
Where $\psi(i,k)=\frac{1}{h_{N_d}}K\left(\frac{(z_{ij}^{\prime }-z_{kj}^{\prime})\beta + (x_{ij}^{\prime }-x_{kj}^{\prime})\delta}{h_{N_d}}\right)$, $K$ is a kernel density function while $h_{N_d}$, is a sequence of bandwidths. To illustrate our identification strategy and for the asymptotic derivation, we use the pairwise spatial difference operator, while for the empirical application and for the Monte Carlo simulations, the fixed effect difference is used.
%for any variable $A$with the observation $k$ in the neighborhood $d$ of $i$ as follows:\begin{equation*}\Delta _{d}A_{ij}=A_{ij}-A_{kj}\end{equation*}

\bigskip For the spatial difference operator $\Delta _{d}$, $\Delta
_{d}y_{2ij}=y_{2ij}-y_{2kj}$ with $k$ an observation in the neighborhood $d$
of $i$. Let $\xi_{ij}\equiv\{ x_{ij}, z_{ij},y_{1ij}^{\ast }>0,\gamma
_{id},\theta _{id} \}$ with $\gamma_{id}=\{ \gamma_{kj} \hspace{0.2cm} with \hspace{0.2cm} k \in \mathcal{N}_{id}\cup \{i\}\}$ and $\theta_{id}=\{ \theta_{kj} \hspace{0.2cm} with \hspace{0.2cm} k \in \mathcal{N}_{id}\cup \{i\}\}.$
\begin{eqnarray}
E[\Delta _{d}y_{2ij}|\xi_{ij}, \xi_{kj} ] &=&E[y_{2ij}-y_{2kj}|\xi_{ij}, \xi_{kj}] \\
&=&E[y_{2ij}|\xi_{ij}]-E[y_{2kj}| \xi_{kj}] \\
&=&x_{ij}^{\prime }\delta +\gamma _{aj}+\gamma _{j}+\rho \lambda
(z_{ij}^{\prime }\beta +\theta _{ja}+\theta _{j}) \\\nonumber
&-&[x_{kj}^{\prime }\delta +\gamma _{j\alpha}+\gamma _{j}+\rho \lambda
(z_{kj}^{\prime }\beta +\theta _{ja}+\theta _{j})]\\
&=&\Delta _{d}x_{ij}^{\prime }\delta +\Delta _{d}\gamma _{j\alpha}+\rho \Delta
_{d}\lambda (z_{ij}^{\prime }\beta +\theta _{ja}+\theta _{j})
\end{eqnarray}%
where $\lambda (c)=\phi (c)/\Phi (c)$ is the inverse Mill's ratio while $\phi (c)$  and $\Phi (c)$ are respectively the density and distribution function of a normal random variable with mean zero and variance 1.

To go from Equation (3) to Equation (4) we use the
linearity of expectation and the mean independence of $y_{2ij}$ and $%
y_{1kj}^{\ast }$ conditional on $\xi_{ij}$, as well as the mean independence of $y_{2kj}$ and $y_{1ij}^{\ast }$
conditional on $\xi_{kj}$, since we have assumed in Assumption I2 that $\varepsilon _{1ij}$ and $\varepsilon _{2ij}$ are $iid$.
The separation of the conditional set, $\xi_{ij}$ and $\xi_{kj}$, is possible because we are working with cross-sectional data. Such separation of the conditional set is not possible for panel data.
Indeed, in the context of panel data with individual effects and sample selection, when the differencing is used to remove the fixed-effects, the conditional set cannot be separated as we have done to move from Equation (3) to (4).
%(AK Note: I am sorry,  I think I know what you are trying to say, but may I ask you to rewrite it?)
For example, \cite{kyriazidou1997estimation} has to impose a ``conditional exchangeability" assumption that is conditioned on the variable related to the two periods used in differencing.  In case of models with censoring, \cite{lee2001first} discusses conditions under which first-difference can be applied, and applies the linear implication of the "conditional exchangeability" assumption.  In a similar context using first difference, \cite{rochina1999new} imposes a joint normality between the difference in the error of the outcome equation and the error in the selections equation in the two time periods.\footnote{See \cite{dustmann2007selection} for a review on selection correction in panel data models.}

Estimating equation (6) presents two challenges for the identification of the parameter of interest $\delta$ and the sample selection parameter $\rho$: the sub-location specific difference $\Delta
_{d}\gamma _{j\alpha}$, and the sample selection term $\rho \Delta
_{d}\lambda (z_{ij}^{\prime }\beta +\theta _{j\alpha}+\theta _{j})$.
As for the sub-location specific difference $\Delta
_{d}\gamma _{j\alpha}$, under Assumption I1 and I2, equation (6) becomes
\begin{equation}
E[\Delta _{d}y_{2ij}|\xi_{ij}, \xi_{kj}]=\Delta _{d}x_{ij}^{\prime }\delta +\rho \Delta _{d}\lambda
(z_{ij}^{\prime }\beta +\theta _{j\alpha}+\theta _{j})  \label{condexp}
\end{equation}%
These assumptions allow us to difference-out the sub-location specific
unobserved effect $\gamma _{j\alpha}$, a strategy that was applied by \cite{duranton2011assessing}.

 As for the sample selection term $\rho \Delta
_{d}\lambda (z_{ij}^{\prime }\beta +\theta _{j\alpha}+\theta _{j})$, we see
that it depends on the unobservable sub-location specific and location effects $%
\theta _{j\alpha}+\theta _{j}$. Because that sample selection term is a nonlinear function, a simple
spatial differencing will not always work unlike the case of $\gamma _{ja}$. Therefore, the following
assumption helps us to deal with this challenge:

\textbf{Assumption I3:}\\
(i) \bigskip {The sub-location specific unobservable selection effect is homogeneous
in a neighborhood of the individual i.e. $\Delta _{d}\theta _{ja}=0$ for $d$
small enough.}\newline

(ii) The changes in the inverse Mill's-Ratio in a
neighborhood of the individual i.e.
\begin{equation}
\frac{\lambda (z_{ij}^{\prime }\beta +\theta _{j\alpha_i}+\theta _{j})-\lambda
(z_{ij}^{\prime }\beta )}{\theta _{j\alpha_i}+\theta _{j}}=\lambda ^{\prime
}(c_{i})=\lambda ^{\prime
}(c_{k})=\frac{\lambda (z_{kj}^{\prime }\beta +\theta _{j\alpha_k}+\theta
_{j})-\lambda (z_{kj}^{\prime }\beta )}{\theta _{j\alpha_k}+\theta _{j}}
\label{assump3}
\end{equation}%
for $i$ and $k$ in a neighborhood $d$ small enough, $\theta _{j\alpha_i}+\theta _{j}$ and $\theta _{j\alpha_k}+\theta _{j}$ both different from 0, $\lambda ^{\prime
}(.)$ is the first derivative the inverse Mill's ratio,  $c_i,$ and $c_k$ are, respectively, in the intervals formed by $[z_{ij}^{\prime }\beta,\hspace{0.2cm} z_{ij}^{\prime }\beta +\theta _{j\alpha_i}+\theta _{j}]$ and $[z_{kj}^{\prime }\beta,\hspace{0.2cm} z_{kj}^{\prime }\beta +\theta _{j\alpha_k}+\theta _{j}]$ such that Equation (\ref{assump3}) holds.

%(AK note: Again, I am sorry but I do not understand what you wanted to say in the part after the equation - did you just wanna describe each bit?)

Assumption I3 (i) is similar to assumption I1. It seems plausible that if that assumption holds for the outcome equation, it will hold true for the selection equation as well.

Assumption I3 (ii) is novel and one of the contributions of this paper.  It assumes that if the exact Taylor approximation is applied on the individual inverse Mill's ratio for individuals $i$ and $k$ in the location $j$, the intermediate points $c_i$ and $c_k$ should be similar.  If the level of nonlinearity of $\lambda (.)$  is low, then the assumption will also hold. In the extreme case of local linearity of the inverse Mill's ratio, the Assumption 3 (ii) perfectly holds.

The combination of assumptions I3 (i) and I3 (ii) implies that  $$\lambda (z_{ij}^{\prime }\beta +\theta _{j\alpha_i}+\theta _{j})-\lambda
(z_{ij}^{\prime }\beta )=\lambda (z_{kj}^{\prime }\beta +\theta _{j\alpha_k}+\theta _{j})-\lambda
(z_{kj}^{\prime }\beta ).$$
 Thus, $\Delta_d\lambda (z_{ij}^{\prime }\beta)=\Delta _{d}\lambda
(z_{ij}^{\prime }\beta +\theta _{j\alpha}+\theta _{j})$

{\begin{thm}
Let us consider the sample selection model presented in Equation \ref{Model}. Under assumptions I1 to I3 the parameters  $\delta$ and $\rho$ are identified.
\end{thm}}

\textbf{Proof of Theorem 1}\\
We have already shown that under the assumptions I1 and I2, we can obtain Equation (\ref{condexp}). Applying the assumption I3, to Equation (\ref{condexp}) leads to the following equation
\begin{equation}  \label{condexpdiff}
E[\Delta_d y_{2ij}|\xi_{ij}, \xi_{kj}] =
\Delta_dx^{\prime }_{ij}\delta + \rho \Delta_d\lambda (z_{ij}^{\prime }\beta).
\end{equation}
Thus, assumptions  I1 to I3 are sufficient for the identification of $\delta $ and $\rho$.\\

We have derived the results using the pairwise spatial difference operator.
However, the identification result  holds for other spatial difference operators as well. In the case of the average or kernel difference operator, the conditioning in equation \ref{condexpdiff} is on $\xi_{kj}$ with $k \in \mathcal{N}_{id}$ for the average difference operator and $k$ is in the full sample for the kernel operator. Note that under assumptions I1 and I3,  any difference of the weighted
average in a neighborhood of the individual will enable us to remove the
sub-location specific effect. The conditional expectation presented in Equation (\ref{condexpdiff}) depends on exogenous observable variables and parameters of interest.

%(AK note: I would suggest we expand this part, especially since we use differencing from the average in the empirical application) GT: I do not have any idea to expand this at this point...

\subsection{Estimation and Asymptotic Properties}

In this section, we present an estimation procedure and derive asymptotic properties of the proposed estimator.
 The estimation procedure involves two-steps. In the first step, probit model is estimated and the inverse Mill's ratio predicted. In the second step, a spatial difference operator differences out both location and the sub-location specific unobserved heterogeneity.
The model is then estimated using an ordinary least square estimator.
When we have a sample of $N$ individuals, the estimation procedure is thus as follows:

\begin{itemize}
\item[] \textbf{Step 1:} Estimate $\beta$ by probit with location
effect $\gamma _{j}$; and calculate  $\hat{\lambda}_i =\lambda (z_{ij}^{\prime }\hat{\beta})$.

\item[] \textbf{Step 2:} Estimate $\delta$ and $\rho$ in the OLS regression
\begin{equation}
\Delta_d y_{2ij}=\Delta_dx^{\prime }_{ij}\delta + \rho \Delta%
_d\lambda (z_{ij}^{\prime }\hat{\beta}) +w_{ikj}.  \label{heck}
\end{equation}
\end{itemize}

 Since we used spatial differencing and $\lambda (z_{ij}^{\prime }\hat{%
\beta})$ is estimated in the first step, a particular structure of the variance-covariance matrix emerges. Therefore, we also need to derive the correct estimator of standard errors $w_{ikj}$ which we will do in section 2.3.

We will now show that the estimator obtained by the above procedure is consistent and asymptotically normal. To derive the asymptotic properties we use similar arguments as those used to derive the asymptotic properties of the clustered standard errors. Specifically, the population size of each location is assumed random and bounded almost surely, and the law of large numbers is applied by letting the number of locations (clusters in case of clustered standard errors) go to infinity.

We consider a generic matrix of spatial difference $\Delta $. The matrix form notation of equation (\ref{heck}) can be expressed in as \footnote{The variables without subscript represent vector or matrices of all observation in the sample.}

\begin{equation}
\Delta y_{2}=\Delta x^{\prime }\delta +\rho \Delta \lambda (z^{\prime }\hat{%
\beta})+\Delta \eta
\label{estiqua}
\end{equation}%
where $\eta _{ij}$ are the same error as in standard sample
selection models.\footnote{We assume the notation that $\lambda(z^{\prime }\hat{\beta})$ is a vector with typical element $\lambda(z_{ij}^{\prime }\hat{\beta})$. } Let us denote $\theta =(\delta ,\rho )^{\prime }$ and $%
W=[x^{\prime },\lambda (z^{\prime }\hat{\beta})]$. The  simplified estimation Equation (\ref{estiqua}) is $$%
\Delta y_{2}=\Delta W\theta +\Delta \eta $$ and OLS estimator of $\theta $ is

\begin{equation}
\hat{\theta}=[ (\Delta W)^{\prime }\Delta W]^{-1}[(\Delta W)^{\prime } \Delta y_{2}]
\label{estim}
\end{equation}

The spatial nature of data implies that an observation $k$ with $n$
neighbours may appear in several pairs. This induces correlation in the error term $%
\Delta \eta $ for all $n$ of these pairs because of the spatial differencing
in the second step of the estimation procedure. As a result, a particular structure of the
covariance matrix emerges, and  we need to take that into account when calculating the standard errors.

To proceed further, we need to introduce assumptions under which the asymptotic properties of our estimator are derived. \\
\textbf{Assumption E1:} The sample is formed if $N$ individuals from the population.\\
$(i)$ We observed $\{ x_{ij}, z_{ij}\}$ independent and identically distributed random variable with $i=1,...,N$ and
$j=1,...,J$. \\ $(ii)$ The number of individuals in a location $j$, $N_j$, is exogenous, random, identically distributed with  $N_j<n_0$ almost surely and $E(N_j)<\infty$; where $n_0$ is a scalar.\\
$(iii)$ The outcomes and the latent variables  are independent across location  i.e. $j_1 \neq j_2$ the variables $y_{2ij_1}\bot y_{2ij_2}$  and  $y^*_{1ij_1}\bot y^*_{1ij_2}$.

An implication of assumption E1 $(i)$ in conjunction with assumption I2 is that $\theta_j$ and $\gamma_j$ are $iid$.
However, $within$ a location $j$, there is a certain level of correlation among individuals which operates through $\theta_{j\alpha_i}$ or $\gamma_{j\alpha_i}$. This means that our assumptions restrict how that within-location individual correlations occurs.
%AK Note: could please see if the above two sentences are okay?

Assumption E1 $(ii)$ restricts the location size to be bounded and implies that the number of locations has to grow to achieve a large sample size in our asymptotic calculation. This assumption is similar to those held in the literature of cluster samples asymptotic and it leads to a ``large number of cluster" asymptotic theory similar to the one discussed in \cite{wooldridge2010econometric}, who assumes fixed cluster size.
%AK Note: could we be more specific where in Wooldridge 2010 - I assume that this is a reference to his book
This assumption corresponds to a specific case of the Assumption 1 in \cite{hansen2019asymptotic}, who allow for different cluster size ranging from fixed to infinite. We have, however, derived the asymptotic of our estimator under the more restrictive condition of Assumption E1 ($ii$).
%AK Note: I would suggest to have a footnote in which we explain briefly what is the specific case in Hansen.GT I have added the explanation in the text.
The reason is that it can be proven that under a joint asymptotic ($N,J \rightarrow \infty$), Assumption 1 is equivalent to assuming that the size of the sample in each location is bounded. If we instead allow for a sequential asymptotic where the number of locations is fixed and the sample size goes to infinity, then there exists at least one location with an infinite number of individuals and the inequality used in the proof of Hansen and Lee (2009)'s  Theorem 1 becomes invalid.

To better illustrate our argument, let us consider the location sample size proposed by \cite{hansen2019asymptotic}: $N_j=N^{\alpha}$ with $0\leq \alpha <1$; we can prove that $1-\alpha = \frac{ln(J)}{ln(N)}$. If we allow for a joint asymptotic, $\alpha$ is not define.
%AK Note: location sample size: is it the sample size of a location j? GT: yes it is.
If on the contrary we assume that the number of locations $J$ is fixed, then, $\alpha$ goes to $1$. In both cases, relaying on Hansen and Lee (2009)'s Assumption 1 seems not enough to warrant the desire asymptotic regularities.

\textbf{Assumption E2:} $z'$  and $W$  are full rank column, with each element having up to its $4^{th}$ moment.\\

{\begin{thm}
 We consider the sample selection model presented in Equation \ref{Model}. Under assumptions I1 to I3, E1 and, E2.\\
 $(i)$ $\hat{\theta} \rightarrow^p \theta$ as $N\rightarrow \infty$\\
 $(ii)$ $\sqrt{N} (\hat{\theta}- \theta)\rightarrow^d \mathcal{N}(0, \Theta)$  with $\Theta= C \Gamma C'$\\
 where $C^{-1}=E((\Delta W_{ij})'\Delta W_{ij}),$  $\Gamma=\rho^2 E[(\Delta W_{ij})'\Omega_{ij} \Delta W_{ij}]+E[(\Delta W_{ij})'\Delta e_{ij}\Delta e_{ij}(\Delta W_{ij})],$\\ and $\Omega_{ij}=[\lambda'(z_{ij}^{\prime }\beta)]^2 z_{ij}^{\prime}V_{\beta} z_{ij}$ taking $V_{\beta}$ as the first step probit variance-covariance matrix.

\end{thm}}

\textbf{Proof of Theorem 2:} In appendix.\\

It is important to notice that the same type of asymptotics should be used in a linear model. In this respect, we complement \cite{duranton2011assessing} who propose a correction for the standard errors, but do not discuss the asymptotic properties of their estimators.
Similarly, \cite{black1999better} and \cite{holmes1996effects} use spatial differencing, but do not account for the fact that differencing will lead to a correlation between pairs where an individual is present.
Our asymptotic derivations do account for the presence of correlation between pairs, and are valid not only for a model with but also without sample selection (in our model, the absence of selection implies $\rho=0$).
They also have important practical implications: the consistency of the estimator requires a large number of locations $\gamma _{j}$, and a small number of individuals in each sub-location $\gamma _{j\alpha}$%.

\subsection{Estimator of Variance }
This section derives a procedure estimate the variance-covariance of the estimator in Equation (\ref{estim}) which has a particular structure arising from $(i)$ spatial differencing and $(ii)$ Heckman's two-step estimation procedure.
%GT: I have changed this introduction as it was incorrect.

We consider $B= \left[ (\Delta W)^{\prime }\Delta W\right]^{-1}$ and
$\Sigma=Var[(\Delta W)^{\prime }\Delta\eta]$ such that the conditional variance-covariance matrix of $\hat{\theta}$ is

\begin{equation*}
Var(\hat{\theta})=B\Sigma B^{\prime }
\end{equation*}

Note that
\begin{equation*}
\Sigma=(\Delta W)^{\prime }Var(\Delta \eta )(\Delta W)
\end{equation*}

This means that we need a consistent estimator of $Var(\Delta \eta )$ to
compute correct standard error for $\hat{\theta}$.

Let us consider that $%
Var(\Delta \eta )=V_{1}+V_{2}$ with
\begin{eqnarray*}
V_{1} &=&\Delta Var(e) \Delta ^{\prime } \\
&=\rho ^{2}&\Delta  R\Delta ^{\prime } \\
&&
\end{eqnarray*}%
where $R$ a diagonal matrix of dimension $N$ (total number of observations),
with $d_{ij}=1-\lambda (z_{ij}^{\prime }\beta )[z_{ij}^{\prime }\beta
+\lambda (z_{ij}^{\prime }\beta )]$ as the diagonal elements.
\begin{eqnarray*}
V_{2} &=&\rho ^{2}\Delta Var\left[ \lambda (z^{\prime }\hat{\beta})-
\lambda (z^{\prime }\beta )\right] \Delta ^{\prime } \\
&=&\rho ^{2}\Delta DzV_{\beta}z^{\prime }D\Delta ^{\prime } \\
\end{eqnarray*}%
where $D$ is the square, diagonal matrix of dimension $N$ with $1-d_{ij}$ as
the diagonal elements; $z$ is the data matrix of selection equation; and $%
V_{p}$ is the variance-covariance estimate from the probit estimation of the
selection equation.

{\begin{thm}
 We consider the sample selection model presented in Equation \ref{Model}. Under assumptions I1 to I3, E1 and, E2.
The variance-covariance estimator of the $\hat{\theta}$ is given by \begin{equation}
V_{twostep}=B(\Delta W)^{\prime }[\hat{V}_{1}+\hat{V}_{2}](\Delta
W)B^{\prime }
\label{var-co2st}
\end{equation}

where $\hat{V}_1=\hat{\rho ^{2}}\Delta \hat{R} \Delta^{\prime }$ and $\hat{%
V}_2=\hat{\rho}^2 \Delta \hat{D} z\hat{V}_{\beta}z^{\prime }\hat{D}\Delta^{\prime } $ with all unknown parameters replaced by their estimates.
Moreover, this is a consistent estimator $Var(\hat{\theta})$.
\end{thm}}

\textbf{Proof of Theorem 3:}\\
The result holds by construction.\\

\section{Monte Carlo Simulation}

In this section we present the results of Monte Carlo simulations to (i) to describe the behavior of the estimator proposed in this paper and (ii) offer empirical guidance for applied research. Regarding the latter, we will pay a close attention to the implication of assumption E1 $(ii)$ according to which it is important to have a large number of locations relative to the number of individuals in the sub-locations. Monte Carlo experiments will offer empirical guidance as to when the number of locations is large enough.

The estimator developed in this paper is referred to as the ``Sub-location Differencing" and it accounts for sub-location specific effect $\gamma _{j\alpha}$. To highlight its features, we compare it to two other estimators. One ignores the presence of both  $\gamma _{j}$ and $\gamma _{j\alpha}$  and applies a simple two step estimator with no spatial differencing - we call it ``No-Differencing" estimator. The other accounts only for the location fixed effects $\gamma _{j}$ and we call it ``Location Differencing" estimator. For each estimator, the mean bias and the coverage rate for the 95\% confidence level test are reported in tables \ref{tab:sim1} to \ref{tab:sim3}.

%AK note: could you see if the above paragraph is okay? GT: I have done Little changes

%AK Note: I know that we talked about it a lot, but going through the text, it does not come across from the main text, unless I missed it. Since we explicitly mention it here, I would suggest that we write about it in Section 2.2. GT: Maybe we can add a section after the theorem 2 to discuss its implications?

The data is obtained using the following data generating process.  We assume that there are $J=20, 30, 100$ non-overlapping locations, each location is divided into $s= 2, 4, 8$ sub-locations. There are $n_j=3, 5, 8, 10$ individuals sharing the same sub-location. The latent variables are $y^*_{1ij} = z_{ij}\beta + \theta_{ijs}+\theta_j+\varepsilon_{1ij}$ and $y^*_{2ij} = x_{ij}\delta +  \gamma_{ijs}+\gamma_j+\varepsilon_{2ij}$, where  $\theta_{ija}=10^{-5}j\times s $ and
$\gamma_{ijs}=5j\times s$ is the sub-location specific
  effect, while $\theta_j=10^{-5} j$ and
  $\gamma_j=10j$ are the location effects;
  for all $i$ and $j$, $x_{ij} \sim \mathcal{N}(0,1)$ , $z_{ij} \sim U(0,1)$ each drawn independently; $\delta=1$, $\beta=0.2$.
The error terms in both equations for all $i$ and $j$ are generated as follows: $\varepsilon_{1ij} \sim \mathcal{N}(0,1)$, $\varepsilon_{2ij}= \rho \varepsilon_{1ij}+v_{ij}$ where $v_{ij} \sim \mathcal{N}(0,1)$  is independent of $\varepsilon_{1ij}$ and $\rho=0.7$.

\begin{table}[htbp]
  \centering
  \caption{Simulation Results with 20 Locations}
  \tiny{\begin{tabular}{|c|c|l|r|r|}
    \toprule
    \multicolumn{1}{|l|}{Numb. Of sub-location } & \multicolumn{1}{l|}{sub-location-size} & Estimators & Mean bias & Coverage rate \\
    \midrule
    \multirow{12}[24]{*}{2} & \multirow{3}[6]{*}{3} & No-differencing & -0.079 & 95.8 \\
\cmidrule{3-5}          &       & Location Differecing & -0.393 & 74.2 \\
\cmidrule{3-5}          &       & Sub-location Differencing & -0.039 & 82.2 \\
\cmidrule{2-5}          & \multirow{3}[6]{*}{5} & No-differencing & -0.344 & 94.5 \\
\cmidrule{3-5}          &       & Location Differecing & -0.953 & 82.1 \\
\cmidrule{3-5}          &       & Sub-location Differencing & 0.011 & 87.0 \\
\cmidrule{2-5}          & \multirow{3}[6]{*}{8} & No-differencing & -0.173 & 94.4 \\
\cmidrule{3-5}          &       & Location Differecing & -1.307 & 90.7 \\
\cmidrule{3-5}          &       & Sub-location Differencing & 0.016 & 79.5 \\
\cmidrule{2-5}          & \multirow{3}[6]{*}{10} & No-differencing & 0.233 & 95.8 \\
\cmidrule{3-5}          &       & Location Differecing & -2.264 & 93.6 \\
\cmidrule{3-5}          &       & Sub-location Differencing & 0.067 & 93.6 \\
    \midrule
    \multirow{12}[24]{*}{4} & \multirow{3}[6]{*}{3} & No-differencing & 0.369 & 95.3 \\
\cmidrule{3-5}          &       & Location Differecing & 0.117 & 88.1 \\
\cmidrule{3-5}          &       & Sub-location Differencing & 0.001 & 75.5 \\
\cmidrule{2-5}          & \multirow{3}[6]{*}{5} & No-differencing & 0.996 & 96.0 \\
\cmidrule{3-5}          &       & Location Differecing & 2.798 & 92.7 \\
\cmidrule{3-5}          &       & Sub-location Differencing & -0.019 & 82.5 \\
\cmidrule{2-5}          & \multirow{3}[6]{*}{8} & No-differencing & 0.085 & 94.5 \\
\cmidrule{3-5}          &       & Location Differecing & 5.824 & 95.2 \\
\cmidrule{3-5}          &       & Sub-location Differencing & -0.035 & 80.5 \\
\cmidrule{2-5}          & \multirow{3}[6]{*}{10} & No-differencing & -0.241 & 95.6 \\
\cmidrule{3-5}          &       & Location Differecing & -2.494 & 96.1 \\
\cmidrule{3-5}          &       & Sub-location Differencing & -0.066 & 88.2 \\
    \midrule
    \multirow{12}[24]{*}{8} & \multirow{3}[6]{*}{3} & No-differencing & 0.833 & 94.7 \\
\cmidrule{3-5}          &       & Location Differecing & -0.310 & 94.9 \\
\cmidrule{3-5}          &       & Sub-location Differencing & -0.005 & 66.3 \\
\cmidrule{2-5}          & \multirow{3}[6]{*}{5} & No-differencing & -0.176 & 93.6 \\
\cmidrule{3-5}          &       & Location Differecing & 1.678 & 96.8 \\
\cmidrule{3-5}          &       & Sub-location Differencing & 0.011 & 76.8 \\
\cmidrule{2-5}          & \multirow{3}[6]{*}{8} & No-differencing & 0.013 & 95.4 \\
\cmidrule{3-5}          &       & Location Differecing & -0.269 & 99.0 \\
\cmidrule{3-5}          &       & Sub-location Differencing & 0.006 & 83.3 \\
\cmidrule{2-5}          & \multirow{3}[6]{*}{10} & No-differencing & 0.271 & 95.2 \\
\cmidrule{3-5}          &       & Location Differecing & -2.573 & 99.3 \\
\cmidrule{3-5}          &       & Sub-location Differencing & -0.016 & 86.1 \\
    \bottomrule
    \end{tabular}}%
  \label{tab:sim1}%
\end{table}%
% Table generated by Excel2LaTeX from sheet 'Sheet1'
\begin{table}[htbp]
  \centering
  \caption{Simulation Results with 30 Locations}
   \tiny{
    \begin{tabular}{|c|c|l|r|r|}
    \toprule
    \multicolumn{1}{|l|}{Numb. Of sub-location } & \multicolumn{1}{l|}{sub-location-size} & Estimators & Mean bias & Coverage rate \\
    \midrule
    \multirow{12}[24]{*}{2} & \multirow{3}[6]{*}{3} & No-differencing & -0.392 & 95.3 \\
\cmidrule{3-5}          &       & Location Differecing & 0.497 & 75.1 \\
\cmidrule{3-5}          &       & Sub-location Differencing & -0.006 & 78.4 \\
\cmidrule{2-5}          & \multirow{3}[6]{*}{5} & No-differencing & 0.095 & 95.2 \\
\cmidrule{3-5}          &       & Location Differecing & -0.965 & 84.5 \\
\cmidrule{3-5}          &       & Sub-location Differencing & 0.007 & 84.8 \\
\cmidrule{2-5}          & \multirow{3}[6]{*}{8} & No-differencing & -0.108 & 94.8 \\
\cmidrule{3-5}          &       & Location Differecing & 1.330 & 92.3 \\
\cmidrule{3-5}          &       & Sub-location Differencing & -0.053 & 85.8 \\
\cmidrule{2-5}          & \multirow{3}[6]{*}{10} & No-differencing & 0.043 & 94.5 \\
\cmidrule{3-5}          &       & Location Differecing & -2.052 & 95.8 \\
\cmidrule{3-5}          &       & Sub-location Differencing & -0.028 & 79.2 \\
    \midrule
    \multirow{12}[24]{*}{4} & \multirow{3}[6]{*}{3} & No-differencing & -0.177 & 95.2 \\
\cmidrule{3-5}          &       & Location Differecing & -1.027 & 89.5 \\
\cmidrule{3-5}          &       & Sub-location Differencing & 0.004 & 71.3 \\
\cmidrule{2-5}          & \multirow{3}[6]{*}{5} & No-differencing & -0.227 & 95.3 \\
\cmidrule{3-5}          &       & Location Differecing & -1.387 & 92.8 \\
\cmidrule{3-5}          &       & Sub-location Differencing & -0.017 & 78.7 \\
\cmidrule{2-5}          & \multirow{3}[6]{*}{8} & No-differencing & 0.424 & 93.6 \\
\cmidrule{3-5}          &       & Location Differecing & -1.437 & 95.0 \\
\cmidrule{3-5}          &       & Sub-location Differencing & 0.012 & 84.2 \\
\cmidrule{2-5}          & \multirow{3}[6]{*}{10} & No-differencing & -0.156 & 95.1 \\
\cmidrule{3-5}          &       & Location Differecing & 0.348 & 96.8 \\
\cmidrule{3-5}          &       & Sub-location Differencing & 0.025 & 78.9 \\
    \midrule
    \multirow{12}[24]{*}{8} & \multirow{3}[6]{*}{3} & No-differencing & 0.031 & 95.0 \\
\cmidrule{3-5}          &       & Location Differecing & -0.200 & 95.0 \\
\cmidrule{3-5}          &       & Sub-location Differencing & 0.010 & 67.4 \\
\cmidrule{2-5}          & \multirow{3}[6]{*}{5} & No-differencing & 0.279 & 94.4 \\
\cmidrule{3-5}          &       & Location Differecing & 2.118 & 97.0 \\
\cmidrule{3-5}          &       & Sub-location Differencing & -0.008 & 73.6 \\
\cmidrule{2-5}          & \multirow{3}[6]{*}{8} & No-differencing & -0.101 & 93.2 \\
\cmidrule{3-5}          &       & Location Differecing & 4.108 & 98.0 \\
\cmidrule{3-5}          &       & Sub-location Differencing & 0.005 & 80.8 \\
\cmidrule{2-5}          & \multirow{3}[6]{*}{10} & No-differencing & -1.369 & 95.6 \\
\cmidrule{3-5}          &       & Location Differecing & -1.866 & 99.5 \\
\cmidrule{3-5}          &       & Sub-location Differencing & -0.041 & 83.6 \\
    \bottomrule
    \end{tabular}}%
  \label{tab:Sim2}%
\end{table}%

\begin{table}[htbp]
  \centering
  \caption{Simulation Results with 100 Locations}
    \tiny{\begin{tabular}{|c|c|l|r|r|}
    \toprule
    \multicolumn{1}{|l|}{Numb. Of sub-location } & \multicolumn{1}{l|}{sub-location-size} & Estimators & Mean bias & Coverage rate \\
    \midrule
    \multirow{12}[24]{*}{2} & \multirow{3}[6]{*}{3} & No-differencing & -0.702 & 94.9 \\
\cmidrule{3-5}          &       & Location Differecing & 1.017 & 74.0 \\
\cmidrule{3-5}          &       & Sub-location Differencing & 0.007 & 63.6 \\
\cmidrule{2-5}          & \multirow{3}[6]{*}{5} & No-differencing & -0.716 & 95.3 \\
\cmidrule{3-5}          &       & Location Differecing & -2.314 & 82.0 \\
\cmidrule{3-5}          &       & Sub-location Differencing & -0.001 & 72.2 \\
\cmidrule{2-5}          & \multirow{3}[6]{*}{8} & No-differencing & -0.089 & 94.6 \\
\cmidrule{3-5}          &       & Location Differecing & -5.000 & 90.5 \\
\cmidrule{3-5}          &       & Sub-location Differencing & -0.005 & 84.2 \\
\cmidrule{2-5}          & \multirow{3}[6]{*}{10} & No-differencing & 0.198 & 94.9 \\
\cmidrule{3-5}          &       & Location Differecing & -0.197 & 96.2 \\
\cmidrule{3-5}          &       & Sub-location Differencing & -0.050 & 85.3 \\
    \midrule
    \multirow{12}[24]{*}{4} & \multirow{3}[6]{*}{3} & No-differencing & -0.096 & 94.5 \\
\cmidrule{3-5}          &       & Location Differecing & 0.858 & 89.1 \\
\cmidrule{3-5}          &       & Sub-location Differencing & -0.001 & 59.4 \\
\cmidrule{2-5}          & \multirow{3}[6]{*}{5} & No-differencing & -0.691 & 94.7 \\
\cmidrule{3-5}          &       & Location Differecing & 0.712 & 89.6 \\
\cmidrule{3-5}          &       & Sub-location Differencing & 0.001 & 64.0 \\
\cmidrule{2-5}          & \multirow{3}[6]{*}{8} & No-differencing & -0.177 & 96.7 \\
\cmidrule{3-5}          &       & Location Differecing & 1.216 & 96.9 \\
\cmidrule{3-5}          &       & Sub-location Differencing & 0.015 & 79.2 \\
\cmidrule{2-5}          & \multirow{3}[6]{*}{10} & No-differencing & -0.384 & 94.0 \\
\cmidrule{3-5}          &       & Location Differecing & -10.139 & 97.9 \\
\cmidrule{3-5}          &       & Sub-location Differencing & -0.075 & 79.6 \\
    \midrule
    \multirow{12}[24]{*}{8} & \multirow{3}[6]{*}{3} & No-differencing & -1.018 & 95.2 \\
\cmidrule{3-5}          &       & Location Differecing & 0.598 & 94.1 \\
\cmidrule{3-5}          &       & Sub-location Differencing & 0.002 & 53.7 \\
\cmidrule{2-5}          & \multirow{3}[6]{*}{5} & No-differencing &   0.050 &  95.1 \\
\cmidrule{3-5}          &       & Location Differecing & 1.400 & 95.7 \\
\cmidrule{3-5}          &       & Sub-location Differencing &   -0.001 & 61.0 \\
\cmidrule{2-5}          & \multirow{3}[6]{*}{8} & No-differencing & \textcolor[rgb]{ .125,  .122,  .118}{ -1.093} & \textcolor[rgb]{ .125,  .122,  .118}{95.7} \\
\cmidrule{3-5}          &       & Location Differecing & \textcolor[rgb]{ .125,  .122,  .118}{ -10.195} & \textcolor[rgb]{ .125,  .122,  .118}{98.1} \\
\cmidrule{3-5}          &       & Sub-location Differencing & \textcolor[rgb]{ .125,  .122,  .118}{ 0.010} & \textcolor[rgb]{ .125,  .122,  .118}{ 75.7} \\
\cmidrule{2-5}          & \multirow{3}[6]{*}{10} & No-differencing & \multicolumn{1}{p{4em}|}{\textcolor[rgb]{ .125,  .122,  .118}{   -0.642}} & \textcolor[rgb]{ .125,  .122,  .118}{ 94.7} \\
\cmidrule{3-5}          &       & Location Differecing & \textcolor[rgb]{ .125,  .122,  .118}{ 6.82} & \textcolor[rgb]{ .125,  .122,  .118}{ 98.9} \\
\cmidrule{3-5}          &       & Sub-location Differencing & \textcolor[rgb]{ .125,  .122,  .118}{0.009} & \textcolor[rgb]{ .125,  .122,  .118}{79.1} \\
    \bottomrule
    \end{tabular}}%
  \label{tab:sim3}%
\end{table}%

%%%%%%%Main Results of the MC simulation%%%%%%%%%%%%%%%%%%%%%%%%%%%%%%%%%%%

We summarize the main results of the simulations in the four points below but in general, the ``Sub-location Differencing" estimator has the smallest mean bias and delivers a conservative coverage rate.\footnote{There is room for improvement concerning our inference strategy. Cluster robust inference is part of a large and growing literature and our work gives some insight as to how diffrencing can be used in cross-sectional data. Future work will investigate the importance of heteroscedasticity, and  small sample procedures such as bootstrap will be used to improve inference.}

\begin{enumerate}
    \item As expected, the ``No-Differencing" estimator has a larger mean bias in the presence of spatial heterogeneity. This result holds for both small and large numbers of locations as well as for few or many individuals having the same sub-location specific unobserved heterogeneity $\mathcal{N}_{id}$.

    \item The mean bias of the ``Sub-location Differencing" estimator is smaller than other estimators. It increases with the number of individuals in the sub-locations, and decreases with the number of locations.  For example,  in a sample of 600 individuals which are spread across 100 locations with 2 sub-locations and 3 individuals in each sub-location, the mean bias is of 0.007. However, for the same sample size but spread across 30 locations with 2 sub-locations and 10 individuals in each sub-locations, the mean bias is $-0.028$.  This result is in line with our asymptotic derivations.
    %AK Note: this item seems to be incorrect: the number of sub-locations is kept fixed at 2 but the text says that it decreases with the number of sub-locations. I am not sure what you had in mind here.  GT: The sentence was not clear and is is surprising that we did get this in our first reading. Thanks.

    \item  For a fixed number of locations, the bias increases with the number of individuals in sub-locations.  The empirical consequence of this results is that our estimator should be applied when the size of sub-locations is small.

    \item The coverage rate of the ``Sub-location Differencing" estimator using the variance-covariance estimator in Equation (\ref{var-co2st}) and the normal asymptotic distribution suggests a conservative coverage.
\end{enumerate}

We now turn to an empirical application of our estimation strategy.

%%%%%%%%%%%%%%%%%%%%%%%%%%%%%%%%%%%%%%%%%%%%%%%%%%%%%%%%%%%%%%%%%%%%%%%%%%%%%%%%%
\section{Empirical Application}

This section shows the empirical importance of spatial differencing methodology  proposed in the previous sections. To illustrate the importance of our estimator, we  ask what determines tax rates set by regional governing bodies.\footnote{There is a large literature which examines a range of factors influencing local tax rates e.g. \cite{charney1983intraurban}, \cite{ashworth1997politicians}, \cite{ross1999sorting},   \cite{charlot2007market,charlot2015does}, \cite{crowley2011does}, \cite{baskaran2014identifying},  \cite{buettner2016yardstick}.} This question opens an important issue of identification since circular causation or omitted variable bias leads to biased and inconsistent estimators. We will use our spatial differencing method to examine the case of changes in the Finish local property tax rate at the turn of the millennium.

Finland consists of 411 municipalities (in 1999) spread across 19 regions which choose property tax rate within the limits set by the central government. In 1999, the central government  decided to raise the lower limit for the year 2000 from 0.2\% to 0.5\%. This change created a probability mass of municipalities at the lower bound: more than half of municipalities have a taxation rate of 0.5\%, making the data sample censored. We investigate what affected municipalities tax rate in the year 2000.

We estimate the parameter of the outcome  equation in the model represented as in Equation (2). Specifically, the outcome variable is the level of general property tax in a municipality $i$ in a region $j$, and explanatory variables include municipality’s $i$ age structure of the population, level of municipality’s income, received subsidies, local income tax rate and a dummy for region $j$ in which the municipality $i$ is located. The selection equation determines whether the municipality sets its general tax rate at the mandatory minimum of 0.5\% or above and contains all the variables which are in the outcome equation except for local income tax rate.

As illustrated in Equation (\ref{lin}), there can be an unobserved sub-location specific effect operating at a finer spatial scale than region $j$, in our case at the level of municipalities which region $j$ consists of. Indeed, municipalities’ tax level can depend not only on its population size, income and subsidies received from the central government, but it can also depend on the level of amenities in the municipality. It is usually difficult to measure them. More importantly, even if we have a few measures of amenities or their proxies, they might not be able to capture all of them, leaving some amenities unobserved. In our case, unobserved amenities can be correlated with municipalities’ population, income level, or the level of subsidies which implies that not controlling for them will render the estimates biased and inconsistent. Therefore, using the fact that two municipalities from the same region sharing borders are neighbor, we use our spatial differencing method to tackle this problem.

We estimate equation (\ref{Model}) with spatial differencing conducted as the difference between municipality $i$ and the average of its neighbours. Columns 1 and 2 present the results without spatial differencing, columns 3-6 with spatial differencing. Estimation is conducted with and without regional dummies respectively, and with two different estimators of standard errors: wild cluster bootstrap, and spatially-adjusted standard errors derived in Section 2.3. Clustering of the standard errors is done at the level of region $j$. Since there are only 19 regions, we use wild clustered bootstrap procedure developed by \cite{cameron2008bootstrap}, which properties were studied by e.g. \cite{davidson2010wild}, \cite{mackinnon2013thirty}, and \cite{mackinnon2017wild}. Specifically, we use a recently developed wild bootstrap package $boottest$ by \cite{roodman2019fast}  implemented in Stata.

%GT: I think that the size of the coefficient is more different in  colums (1) (2) compare to other columns.
We begin with discussing the results with spatial differencing: columns 3-6,  Columns 3 and 4 present the results when we use spatial differencing with the standard errors calculated using the formula derived in Section 2.3. There is only one significant variable: the share of population older than 75 in column 4. This is not surprising, since our estimator of variance-covariance matrix is an asymptotic estimator, while the estimation is done on a sample with small number of clusters (nineteen regions). Therefore we use wild-bootstrap procedure which is known to be suitable for a small number of clusters. The results using wild-bootstrapping are shown in columns 5 and 6 and we see a considerable increase in the number of statistically significant results.

The comparison of columns 5 and 6 with columns 1 and 2 reveals the importance of spatial differencing. Controlling for the sub-location specific unobserved effect $\gamma _{j\alpha}$ by spatial differencing  renders four variables statistically significant: share of population younger than 15, share of population older than 75, government grants, and income tax rate. This is in contrast to columns 1 and 2 in which income tax rate is the only significant variable. Not controlling for $\gamma _{j\alpha}$ leads to the omitted variable bias which, apart from rendering the estimates inconsistent, inflates the standard errors and makes the estimates mostly insignificant. Spatial differencing controls for this omitted variable bias, which means that they do not 'end up' in the error term and do not inflate the standard errors. In addition to comparing estimates with and without spatial differencing, it is also instructive to compare columns 5 and 6: spatial differencing with and without regional dummies. We see that controlling for the regional unobserved effect $\gamma_{j}$ does not help to control for sub-location specific effects $\gamma _{j\alpha}$. Indeed, the magnitude and the statistical significance of the estimates change very little, and even the one variable that looses its statistical significance after including regional dummies (municipality's income) is only marginally significant without these dummies.

Overall, our empirical analysis shows that controlling for the unobserved municipality effects matters. Estimations which control for spatial unobserved effects only at the regional suggest that the income tax rate is the only determinant of general tax rate set by the municipalities. However, after controlling for the sub-location specific unobserved effects, the tax rate depends not only on the income tax rate but also on the age composition of their population - the share of young as well as the share of elderly population. These results thus indicate that spatial differencing is an important tool to deal with omitted variable bias which often plagues empirical studies on local taxation.

\begin{table}[htbp]
  \centering
  \caption{Determinants of Municipality Taxation Rate}
   \tiny{ \begin{tabular}{lcccccc}
    \toprule
\multicolumn{1}{c|}{\multirow{4}[1]{*}{ }} & \multicolumn{2}{c|}{No Spatial Differencing } & \multicolumn{4}{c|}{Spatial Differencing } \\
\cmidrule{2-7}    \multicolumn{1}{c|}{} & \multicolumn{2}{c|}{Wild bootstrap} & \multicolumn{2}{c|}{Spatially adj se} & \multicolumn{2}{c|}{Wild bootstrap} \\
\cmidrule{2-7}    \multicolumn{1}{c|}{} & \multicolumn{1}{c|}{\multirow{2}[2]{*}{No Reg. Dummies}} & \multicolumn{1}{c|}{\multirow{2}[2]{*}{Regional Dummies}} & \multicolumn{1}{r|}{\multirow{2}[2]{*}{No Reg. Dummies}} & \multicolumn{1}{c|}{\multirow{2}[2]{*}{Regional Dummies}} & \multicolumn{1}{c|}{\multirow{2}[2]{*}{No Reg. Dummies}} & \multicolumn{1}{c|}{\multirow{2}[2]{*}{Regional Dummies}} \\
    \multicolumn{1}{c|}{} & \multicolumn{1}{c|}{} & \multicolumn{1}{c|}{} & \multicolumn{1}{c|}{} & \multicolumn{1}{c|}{} & \multicolumn{1}{c|}{} & \multicolumn{1}{c|}{} \\
    \midrule
    \multicolumn{1}{c}{ } & (1)    & (2)   & (3)    & \multicolumn{1}{c}{(4)} & \multicolumn{1}{c}{(5)} & \multicolumn{1}{c|}{(6)} \\
    \midrule
    Population  & -1.897 & -0.7047 & 0.1346 & 0.2448 & 0.1346 & 0.2448 \\
          & \multicolumn{1}{c}{[-1.420]} & \multicolumn{1}{c}{[-1.0432]   } & \multicolumn{1}{c}{[0.0489]} & \multicolumn{1}{c}{[0.2656]} & \multicolumn{1}{c}{[0.507]} & \multicolumn{1}{c}{[0.8793]} \\
   Share pop.<15  & 0.009 & -0.0015 & -0.0116 & -0.0113 & \multicolumn{1}{c}{-0.0116**} & \multicolumn{1}{c}{-0.0113**} \\
          & \multicolumn{1}{c}{[0.492]} & \multicolumn{1}{c}{[-0.0996]   } & \multicolumn{1}{c}{[-0.0933]} & \multicolumn{1}{c}{[-0.1373]} & \multicolumn{1}{c}{[-2.536]} & \multicolumn{1}{c}{[-2.3921]} \\
    Share 61<pop.<74 & -0.0055 & -0.0054 & -0.0089 & -0.0092 & -0.0089 & -0.0092 \\
          & \multicolumn{1}{c}{[-0.923]} & \multicolumn{1}{c}{[-0.343]   } & \multicolumn{1}{c}{[-0.1106]} & \multicolumn{1}{c}{[-0.0721]} & \multicolumn{1}{c}{[-1.495]} & \multicolumn{1}{c}{[-1.4902]} \\
    Share pop.>75 & 0.0211 & 0.0049 & -0.0145 & \multicolumn{1}{c}{-0.0140**} & \multicolumn{1}{c}{-0.0145*} & \multicolumn{1}{c}{-0.014*} \\
          & \multicolumn{1}{c}{[1.027]} & \multicolumn{1}{c}{[0.4688]   } & \multicolumn{1}{c}{[-0.1575]} & \multicolumn{1}{c}{[-1.8771]} & \multicolumn{1}{c}{[-1.794]} & \multicolumn{1}{c}{[-1.6746]} \\
    Income & 2.40E-07 & 6.48E-06 & 1.23E-05 & 0.00001 & \multicolumn{1}{c}{1.23E-05*} & 1.1E-05 \\
          & \multicolumn{1}{c}{[0.047]} & \multicolumn{1}{c}{[0.8092]   } & \multicolumn{1}{c}{[0.0234]} & \multicolumn{1}{c}{[0.0213]} & \multicolumn{1}{c}{[1.924]} & \multicolumn{1}{c}{[1.6688]} \\
    Gov. grant  & -1.7E-05 & 1.8E-05 & -1.78E-05 & -2.16E-05 & -1.78E-05 & -2.16E-05 \\
          & \multicolumn{1}{c}{[-0.868]} & \multicolumn{1}{c}{[0.2555]   } & \multicolumn{1}{c}{[-0.0458]} & \multicolumn{1}{c}{[-0.0925]} & \multicolumn{1}{c}{[-0.483]} & \multicolumn{1}{c}{[-0.5666]} \\
    Income tax rate & \multicolumn{1}{c}{0.0350***} & \multicolumn{1}{c}{0.0396**   } & 0.0482 & 0.0453 & \multicolumn{1}{c}{0.0482***} & \multicolumn{1}{c}{0.0453***} \\
          & \multicolumn{1}{c}{[3.383]} & \multicolumn{1}{c}{[3.547]   } & \multicolumn{1}{c}{[0.2775]} & \multicolumn{1}{c}{[0.2965]} & \multicolumn{1}{c}{[3.362]} & \multicolumn{1}{c}{[3.088]} \\
    Inverse Mill's ratio & 0.3028 & 0.1731 & 0.0159 & 0.0034 & 0.0159 & 0.0034 \\
          & \multicolumn{1}{c}{[1.505]} & \multicolumn{1}{c}{[1.1468]   } & \multicolumn{1}{c}{[0.0346]} & \multicolumn{1}{c}{[0.0298]} & \multicolumn{1}{c}{[0.854]} & \multicolumn{1}{c}{[0.1525]} \\
    Constant & -0.5327 & -0.3601 & 0.0111 & -0.006 & 0.0111 & -0.006 \\
          & \multicolumn{1}{c}{[-0.713]} & \multicolumn{1}{c}{[-0.598]   } & \multicolumn{1}{c}{[1.540]} & \multicolumn{1}{c}{[-0.230]} & \multicolumn{1}{c}{[1.540]} & \multicolumn{1}{c}{[-0.199]} \\
    Observations & 403   & 403   & 273   & 273   & 273   & 273 \\
    Regional Dummies & \multicolumn{1}{c}{NO} & \multicolumn{1}{c}{YES   } & \multicolumn{1}{c}{NO} & \multicolumn{1}{c}{YES} & \multicolumn{1}{c}{NO} & \multicolumn{1}{c}{YES} \\
    Number of Dummies & 19    & 19    & 19    & 19    & 19    & 19 \\
    R-squared & 0.197 & 0.271 & 0.248 & 0.279 & 0.248 & 0.279 \\
    \bottomrule
\multicolumn{7}{p{18.5cm}}{ \footnotesize{\textbf{Source}: see text; Note: t-statistics in brackets, *** p<0.01, ** p<0.05, * p<0.1. The table reports regression results with municipality taxation rate as the dependent variable. Selection equation excludes local income tax rate. Wild bootstrapping with 999 interactions was used except for column 4 which reports t-statistics calculated with the standard errors adjusted for spatial differencing.}} \\
    \end{tabular}}%
  \label{tabTax}%
\end{table}%

\section{Conclusion}

This paper has investigated a sample selection model with unobserved
heterogeneity at a very fine location level. It proposes spatial differencing as an alternative identification strategy when instrumental variable and/or a panel data are not available.  We discuss the assumptions under which the parameters of the model are identified. The estimation of the parameters is done using the classic Heckman's two-step estimation procedure. The differecing and the two-step procedure lead to  a novel estimator with properties that are also relevant for spatial differencing in linear models. To understand the behavior of the new estimator, we derive a cluster asymptotic of the estimator. The derivation reveals two important implications for its empirical implementation: $(i)$ the number of clusters needs to be large for inference to be based on normal distribution. ($ii$) each cluster should have a bounded number of individuals.

Monte Carlo experiments show that accounting for sub-location specific heterogeneity is crucial for identification. It also confirms the estimator's properties derived in our asymptotic.  In particular, the estimator performs better with the increasing number of locations, and fewer individuals in sub-locations. In addition, ignoring sub-locations and applying spatial differencing only to more aggregate geographical units subsuming sub-locations, the mean bias is larger. The coverage rate of the test based on the corrected standard error has an empirical coverage lower that the  theoretical one.

In the empirical application which looked at the determinants of municipal tax rate, we show that using spatial differencing in combination with cluster wild-bootstrap inference tools can be extremely useful. Indeed, the new estimator reveals several determinants of the municipal tax rate that would have been missed otherwise. The development of a bootstrap appropriate sample selection models is left for future research.

\newpage %\bibliographystyle{plain}

\section*{Appendix}

\textbf{Proof of Theorem 2:}\\
The proof is written conditional on the set of number of individuals in the locations. Thus, when $E(N_j)$ is used, it can be considered as a constant.

 The substitution of the true value of $\Delta y_{2}$  in  equation (\ref{estim})
yields the following equality \begin{equation*}
\hat{\theta}=\theta+ [ (\Delta W)^{\prime }\Delta W]^{-1}[(\Delta W)' \Delta \eta]
\end{equation*}

Let us assume that $y_{2ij}=x_{ij}^{\prime }\delta +\gamma _{j\alpha}+\gamma
_{j}+ \rho \lambda (z_{ij}^{\prime }\beta +\theta _{j\alpha_k}+\theta _{j})+ e_{ij} $ with $E(e_{ij}| \xi_{ij})=0$. Thus, $\Delta y_{2ij}=\Delta x_{ij}^{\prime }\delta +\rho \Delta \lambda (z_{ij}^{\prime }\beta +\theta _{j\alpha}+\theta _{j})+ \Delta e_{ij}$. Under the identification assumptions I1 to I3 have
$$\Delta y_{2ij}=\Delta x_{ij}^{\prime }\delta +\rho \Delta \lambda (z_{ij}^{\prime }\beta )+ \Delta e_{ij}.$$

The second step regression equation is equivalent to

$$\Delta y_{2ij}=\Delta x_{ij}^{\prime }\delta +\rho \tilde{\Delta} \lambda (z_{ij}^{\prime }\beta )+ \Delta [\rho (\lambda (z_{ij}^{\prime }\beta)- \lambda (z_{ij}^{\prime }\hat{\beta}))+ e_{ij}]$$

$\hat{\beta}$ is estimated by maximum likelihood  probit in the first step with variance-covariance matrix $V_{\beta}$. Given that $\lambda (.)$ is twice differentiable, the continuous mapping theorem implies that  $\lambda (z_{ij}^{\prime }\beta)- \lambda (z_{ij}^{\prime }\hat{\beta})$ goes to zero in probability and is asymptotically normal. If we assume that $N_1$ is the full sample while $N$ is the selected sample.\footnote{ We assume that $N/N_1 \rightarrow 1$.} We, therefore, have
\begin{equation}\label{probit}
  \sqrt{N_1}( \lambda (z_{ij}^{\prime }\beta)- \lambda (z_{ij}^{\prime }\hat{\beta}))\rightarrow^d \mathcal{N}(0, \Omega_{ij})
\end{equation}

where $\Omega_{ij}= [\lambda'(z_{ij}^{\prime }\beta)]^2 z_{ij}^{\prime}V_{\beta} z_{ij}$.\\

We are interested in the limiting distribution of $\sqrt{N}( \hat{\theta}- \theta)$.
\begin{eqnarray*}
% \nonumber to remove numbering (before each equation)
  \sqrt{N}( \hat{\theta}- \theta) &=& N[ (\Delta W)^{\prime }\Delta W]^{-1}\frac{1}{\sqrt{N}}[(\Delta W)' \Delta \eta] \\
   &=& \left[\frac{\sum_{i=1}^{N}(\Delta W_i)'\Delta W_i}{N}\right]^{-1}\frac{1}{\sqrt{N}}\sum_{i=1}^{N}(\Delta W_i)'\Delta \eta_i  \\
\end{eqnarray*}

 While $W_i$ are $iid$, $\Delta W_i$ are not independent because an individual is allowed to appear in many pairs. We will therefore have to use LLN and CLT for non-independent random variables.
The dependence structure is driven by the operator $\Delta$. If $\Delta$ is such that each individual appear only in one pair then the classical CLT and LLN could be applied. However, if individuals are allowed to appear in several pairs, then we need to apply CLT and LLN accounting for correlation.

\begin{eqnarray*}
% \nonumber to remove numbering (before each equation)
  \frac{\sum_{i=1}^{N}(\Delta W_i)'\Delta W_i}{N} &=& \frac{\sum_{j=1}^J\sum_{k=1}^{N_j}(\Delta W_{kj})'\Delta W_{kj}}{N} \\
   &=& \frac{1}{J}\sum_{j=1}^J\frac{1}{E(N_j)}\sum_{k=1}^{N_j}(\Delta W_{kj})'\Delta W_{kj}
\end{eqnarray*}

Let us consider $Y_j=\frac{1}{E(N_j)}\sum_{k=1}^{N_j}(\Delta W_{kj})'\Delta W_{kj}$, these variables are $iid$, moreover, note that $N=N_1+N_2+....+N_J=JE(N_j)$.
Under assumption E1, all locations have a bounded maximum capacity of $N_j<n_0$ with $n_0$ a scalar.
Under the assumption that all second moments of the variables in  $W$ exist (Assumption E2).\\ $H_J=\frac{1}{J}\sum_{j=1}^J\frac{1}{E(N_j)}\sum_{k=1}^{N_j}Y_j$ is a matrix. Thus, the law of large number apply to it if and only if it applies to all is elements.  Let $a_j$ be a typical element of the matrix $\frac{1}{E(N_j)}\sum_{k=1}^{N_j}Y_j$. Let $t$ and $m$ be two variables from the set of variables forming $W$. For example, we can consider $t=x_1$ the first column of the random variable $x$. If $t\neq m$ then,

\begin{eqnarray}
% \nonumber to remove numbering (before each equation)
  E|a_j| & \leq & \frac{1}{E(N_j)}\sum_{k=1}^{N_j}E|\Delta t_{kj}\Delta m_{kj}| \\
   &=&  \frac{1}{E(N_j)}\sum_{k=1}^{N_j}E| (t_{kj}-t_{ij})( m_{kj}-m_{ij})|  \\
   &\leq & \frac{4}{E(N_j)}\sum_{k=1}^{N_j}E|t_{kj}m_{kj}| \\
   &\leq &  \frac{4}{E(N_j)}\sum_{k=1}^{N_j}\sqrt{E(|t_{kj}|^2) E(|m_{kj}|^2)}\\
   &\leq & M_0
\end{eqnarray}

with $M_0$ a constant. \\

The result is obtained by using successively  the triangular inequality, the identical distribution of variable in $W$, the Cauchy-Schwars's inequality and the existence of moment up-to its fourth (which means that the second moment exists).
If $t=m$ we have,
\begin{eqnarray}
% \nonumber to remove numbering (before each equation)
  E|a_j| & \leq & \frac{1}{E(N_j)}\sum_{k=1}^{N_j}E|\Delta t_{kj}\Delta t_{kj}| \\
   &=&  \frac{1}{E(N_j)}\sum_{k=1}^{N_j}E| (t_{kj}-t_{ij})^2|  \\
   &\leq & \frac{2}{E(N_j)}\sum_{k=1}^{N_j}E|t_{kj}t_{ij}|+ E(t_{kj}^2) \\
   &\leq &  \frac{1}{E(N_j)}\sum_{k=1}^{N_j}(E|t_{kj}|)^2+ E(t_{kj}^2)\\
   &\leq & M_0
\end{eqnarray}

Thus, the LLN implies that $$\frac{\sum_{i=1}^{N}(\Delta W_i)'\Delta W_i}{N}\rightarrow^p E((\Delta W_{ij})'\Delta W_{ij})=C^{-1}.$$

  We can also show that
   \begin{eqnarray*}
   % \nonumber to remove numbering (before each equation)
\frac{1}{\sqrt{N}}\sum_{i=1}^{N}(\Delta W_i)'\Delta \eta_i      &=&  \frac{\rho}{\sqrt{N}}\sum_{i=1}^{N}(\Delta W_i)'\Delta ( \lambda (z_{ij}^{\prime }\beta)- \lambda (z_{ij}^{\prime }\hat{\beta})) \\
      &+&  \frac{1}{\sqrt{N}}\sum_{i=1}^{N}(\Delta W_i)'\Delta e_{ij}.
   \end{eqnarray*}

We consider $\Lambda_j=\sum_{k=1}^{N_j}(\Delta W_{kj})'\Delta ( \lambda (z_{kj}^{\prime }\beta)- \lambda (z_{kj}^{\prime }\hat{\beta}))$ and $E_j= \sum_{k=1}^{N_j}(\Delta W_{kj})'\Delta e_{kj}$. Conditional on $\hat{\beta}$, $\Lambda_j$ are $iid$ random variables; $E_j$ are too. We assume that the  number of individuals in a group is $iid$ with finite mean $E(N_j)$.\footnote{The application of the LLN implies for consistency reason that $ N/J\rightarrow^p E(N_j).$ Thus $JE(N_j) \approx N.$} Given all locations are assumed to be disjoint,
\begin{eqnarray*}
% \nonumber to remove numbering (before each equation)
  \frac{1}{\sqrt{N}}\sum_{i=1}^{N}(\Delta W_i)'\Delta \eta_i &=& \frac{\rho}{\sqrt{N}}\sum_{j=1}^{J} \Lambda_j \\
   &+& \frac{1}{\sqrt{N}}\sum_{j=1}^{J}E_j.
\end{eqnarray*}

We have $E(E_j)=0$ for each $j.$ Moreover, \begin{eqnarray*}
  % \nonumber to remove numbering (before each equation)
  Var(E_j) &=& E[\sum_{k=1}^{N_j}(\Delta W_{kj})'\Delta e_{kj}(\sum_{k=1}^{N_j}(\Delta W_{kj})'\Delta e_{kj})'] \\
                                              &=& E[\sum_{k=1}^{N_j}(\Delta W_{kj})'\Delta e_{kj}\Delta e_{kj}(\Delta W_{kj})] \\
                                              &=& E(N_j) E[(\Delta W_{kj})'\Delta e_{kj}\Delta e_{kj}(\Delta W_{kj})].
   \end{eqnarray*}
Under Assumption E2, $Var(E_j)$ is finite, because all variables have up-to the fourth moments. Indeed, if we consider a typical element of $E[(\Delta W_{kj})'\Delta e_{kj}\Delta e_{kj}(\Delta W_{kj})]$, form by the variables $t$ and $m$,

\begin{eqnarray*}
% \nonumber to remove numbering (before each equation)
  E[(t_{kj}-t_{ij})( m_{kj}-m_{ij})\Delta e_{kj}(\Delta W_{kj})] &\leq & 4E[t_{kj}m_{kj}(\Delta e_{kj})^2] \\
   &\leq & 4E|t_{kj}m_{kj}\Delta e_{kj}^2| \\
   &\leq & 4\sqrt[4]{E(|t_{kj}|^4)E[(\Delta e_{kj})^2] E(|m_{kj}|^4) E[(\Delta e_{kj})^2]} \\
   &\leq & M_0
\end{eqnarray*}

It should be noted that $ E[(\Delta e_{kj})^2]=2E( e_{kj}^2)<\infty$.\\

Similarly, we can show that $E(\Lambda_j)=0$, and
\begin{eqnarray}
% \nonumber to remove numbering (before each equation)
 \nonumber Var(\Lambda_j) &=& E[(\sum_{k=1}^{N_j}(\Delta W_{kj})'\Delta ( \lambda (z_{kj}^{\prime }\beta)- \lambda (z_{kj}^{\prime }\hat{\beta})))(\sum_{k=1}^{N_j}(\Delta W_{kj})'\Delta ( \lambda (z_{kj}^{\prime }\beta)- \lambda (z_{kj}^{\prime }\hat{\beta})))'] \\
    &=& E[\sum_{k=1}^{N_j}(\Delta W_{kj})'\Delta ( \lambda (z_{kj}^{\prime }\beta)- \lambda (z_{kj}^{\prime }\hat{\beta}))(\Delta W_{kj})'\Delta ( \lambda (z_{kj}^{\prime }\beta)- \lambda (z_{kj}^{\prime }\hat{\beta}))'] \\
   &=& E(N_j) E((\Delta W_{kj})'\Delta ( \lambda (z_{kj}^{\prime }\beta)- \lambda (z_{kj}^{\prime }\hat{\beta}))\Delta ( \lambda (z_{kj}^{\prime }\beta)- \lambda (z_{kj}^{\prime }\hat{\beta}))(\Delta W_{kj}))\\
   &=& E(N_j) E[(\Delta W_{kj})'\Omega_{kj}(\Delta W_{kj})]
\end{eqnarray}

We need to show that $ E[(\Delta W_{kj})'\Omega_{kj}(\Delta W_{kj})]$, with $\Omega_{kj}= [\lambda'(z_{kj}^{\prime }\beta)]^2 z_{kj}^{\prime}V_{\beta} z_{kj}$ is finite.   A typical element of this matrix is given by, $ E[(t_{kj}-t_{ij})\Omega_{kj}( m_{kj}-m_{ij})]$. We can show the following using a Cauchy-Schwarz's inequality.\\

\begin{eqnarray}
% \nonumber to remove numbering (before each equation)
  E[(t_{kj}-t_{ij})\Omega_{kj}( m_{kj}-m_{ij})] &\leq & 4 E[t_{kj}m_{kj}\Omega_{kj}]  \\
   &\leq & 4 E[|t_{kj}m_{kj}\Omega_{kj}|]  \\
\nonumber   &\leq&  4 \sqrt[4]{E(|t_{kj}|^4) (E([\lambda'(z_{kj}^{\prime }\beta)]^2 z_{kj}^{\prime}V_{\beta} z_{kj}))^2 E(|m_{kj}|^4) }
\end{eqnarray}

It remains to be proofed that $E([\lambda'(z_{kj}^{\prime }\beta)]^2 z_{kj}^{\prime}V_{\beta} z_{kj})<\infty$.  The application of the Cauchy-Schwarz's inequality implies,
\begin{eqnarray}
% \nonumber to remove numbering (before each equation)
   E([\lambda'(z_{kj}^{\prime }\beta)]^2 z_{kj}^{\prime}V_{\beta} z_{kj}) & \leq & \sqrt{ E([\lambda'(z_{kj}^{\prime }\beta)]^4 E[(z_{kj}^{\prime}V_{\beta} z_{kj})^2]} \\
     &\leq &  \sqrt{  E[(z_{kj}^{\prime}V_{\beta} z_{kj})^2]} <\infty
\end{eqnarray}
This follows from noting that $|\lambda'(.)| \leq 1$  and the elements of $z$ have up-to their fourth moments.\\

The moment of a typical element $ E[(t_{kj}-t_{ij})\Omega_{kj}( m_{kj}-m_{ij})]<\infty.$

This proof that the variance is finite.\\

 It is important to notice that conditional $W$, $\sum_{i=1}^{N}(\Delta W_i)'\Delta ( \lambda (z_{ij}^{\prime }\beta)- \lambda (z_{ij}^{\prime }\hat{\beta}))$ and   $\sum_{i=1}^{N}(\Delta W_i)'\Delta e_{ij}$ are independent random variables.  Therefore,

  \begin{equation}\label{nor}
\frac{1}{\sqrt{N}}\sum_{i=1}^{N}(\Delta W_i)'\Delta \eta_i \rightarrow^d \mathcal{N}(0, \Gamma),
  \end{equation}
 where $\Gamma=\rho^2 E[(\Delta W_{ij})'\Omega_{ij} \Delta W_{ij}]+E[(\Delta W_{ij})'\Delta e_{ij}\Delta e_{ij}(\Delta W_{ij})].$
   \begin{equation}\label{normd}
  \sqrt{N}( \hat{\theta}- \theta)\rightarrow^d \mathcal{N}(0, \Theta)
   \end{equation}
 with $\Theta= C \Gamma C'$. This proves the asymptotic normality of our two step estimator. \\

 We have proven that under assumptions I1, I2, I3, E1 and E2,  $$\frac{\sum_{i=1}^{N}(\Delta W_i)'\Delta W_i}{N}\rightarrow^p E((\Delta W_1)'\Delta W_1)=C^{-1}.$$ Using similar arguments  we can show that
  $$\frac{\sum_{i=1}^{N}(\Delta W_i)'\Delta \eta_i}{N}\rightarrow^p E((\Delta W_1)'\Delta \eta)=0.$$
  Which means that $\hat{\theta}$ is a consistent estimator of $\theta.$
 We have  proven the estimator is both consistent and asymptotically normal. This ends the proof of Theorem 2.\\

\newpage

\bibliographystyle{econometrica}
\bibliography{Bib_Spatial}

\end{document}